\documentclass[aps,prx,twocolumn,superscriptaddress,linenumbers]{revtex4}
\usepackage{amsmath}
\usepackage{amssymb}
\usepackage{graphicx}
\usepackage{mathtools}
\usepackage{relsize}
\usepackage{lmodern}
\usepackage{slantsc}
\usepackage{scalefnt}
\usepackage{subfigure}
\usepackage{dsfont}
\usepackage{ifpdf}
\usepackage{pgffor}
\usepackage{verbatim}
\usepackage{tikz}
\usetikzlibrary{ }
\usepackage{color}
\usetikzlibrary{arrows,matrix,calc,scopes,decorations.markings}
\allowdisplaybreaks[1]

\newcommand{\be}{\begin{equation}}
\newcommand{\ee}{\end{equation}}
\newcommand{\bse}{\begin{subequations}}
\newcommand{\ese}{\end{subequations}}
\newcommand{\ket}[1]{|{#1}\rangle}

\newcommand{\Z}{\mathbb{Z}}
\newcommand{\ii}{\mathrm{i}}

\newcommand{\bpm}{\begin{pmatrix}}
\newcommand{\epm}{\end{pmatrix}}
\newcommand{\bmm}{\begin{matrix}}
\newcommand{\emm}{\end{matrix}}

\makeatletter
\newcommand*{\Relbarfill@}{\arrowfill@\Relbar\Relbar\Relbar}
\newcommand*{\xeq}[2][]{\ext@arrow 0055\Relbarfill@{#1}{#2}}
\newcommand{\x}{\times}

\newcommand{\state}[2]{\bigl\vert \begin{smallmatrix} #1\\ #2 \end{smallmatrix}\bigr\rangle}
\makeatother

\tikzset{->-/.style={decoration={
        markings,
        mark=at position .5 with {\arrow{>}}},postaction={decorate}}}
\tikzset{-<-/.style={decoration={
        markings,
        mark=at position .5 with {\arrow{<}}},postaction={decorate}}}

\usetikzlibrary{decorations.markings}
\usetikzlibrary{decorations.pathmorphing,positioning}
\tikzset{>=latex}
\tikzset{snake it/.style={decorate, decoration={snake,amplitude=0.15mm,segment length=1mm}}}
\tikzset{->-/.style={decoration={
			 markings,
			 mark=at position .55 with {\arrow{>}}},postaction={decorate}}}

\tikzset{-<-/.style={decoration={
			 markings,
			 mark=at position .55 with {\arrow{<}}},postaction={decorate}}}

\newcommand{\AnomalyPic}{
\begin{tikzpicture}[scale=0.7]
\draw [lightgray, fill] (3.4,9.8) ..controls (3.2,9.861) and (2.985,9.846) .. (2.6,9.8)
			 ..controls (2.215,9.754) and (1.659,9.676) .. (1.6,9.4)
			 ..controls (1.541,9.124) and (1.98,8.651) .. (2.2,8.6)
			 ..controls (2.42,8.549) and (2.42,8.92) .. (2.6,9.)
			 ..controls (2.78,9.08) and (3.141,8.871) .. (3.4,8.9)
			 ..controls (3.659,8.929) and (3.815,9.198) .. (3.8,9.4)
			 ..controls (3.785,9.602) and (3.6,9.739) .. (3.4,9.8);
\draw [lightgray, fill] (9.2,9.8) ..controls (8.947,9.867) and (8.413,9.627) .. (8.,9.6)
			 ..controls (7.587,9.573) and (7.293,9.76) .. (7.2,9.6)
			 ..controls (7.107,9.44) and (7.213,8.933) .. (7.4,8.8)
			 ..controls (7.587,8.667) and (7.853,8.907) .. (8.2,9.)
			 ..controls (8.547,9.093) and (8.973,9.04) .. (9.2,9.2)
			 ..controls (9.427,9.36) and (9.453,9.733) .. (9.2,9.8);
\draw [very thick] (3.4,7.3) ..controls (3.2,7.361) and (2.985,7.346) .. (2.6,7.3)
			 ..controls (2.215,7.254) and (1.659,7.176) .. (1.6,6.9)
			 ..controls (1.541,6.624) and (1.98,6.151) .. (2.2,6.1)
			 ..controls (2.42,6.049) and (2.42,6.42) .. (2.6,6.5)
			 ..controls (2.78,6.58) and (3.141,6.371) .. (3.4,6.4)
			 ..controls (3.659,6.429) and (3.815,6.698) .. (3.8,6.9)
			 ..controls (3.785,7.102) and (3.6,7.239) .. (3.4,7.3);
\draw [very thick] (8.,7.1) ..controls (7.587,7.073) and (7.293,7.26) .. (7.2,7.1)
			 ..controls (7.107,6.94) and (7.213,6.433) .. (7.4,6.3)
			 ..controls (7.587,6.167) and (7.853,6.407) .. (8.2,6.5)
			 ..controls (8.547,6.593) and (8.973,6.54) .. (9.2,6.7)
			 ..controls (9.427,6.86) and (9.453,7.233) .. (9.2,7.3)
			 ..controls (8.947,7.367) and (8.413,7.127) .. (8.,7.1);
\draw [very thick] (3.4,9.8) ..controls (3.2,9.861) and (2.985,9.846) .. (2.6,9.8)
			 ..controls (2.215,9.754) and (1.659,9.676) .. (1.6,9.4)
			 ..controls (1.541,9.124) and (1.98,8.651) .. (2.2,8.6)
			 ..controls (2.42,8.549) and (2.42,8.92) .. (2.6,9.)
			 ..controls (2.78,9.08) and (3.141,8.871) .. (3.4,8.9)
			 ..controls (3.659,8.929) and (3.815,9.198) .. (3.8,9.4)
			 ..controls (3.785,9.602) and (3.6,9.739) .. (3.4,9.8);
\draw [very thick] (8.,9.6) ..controls (7.587,9.573) and (7.293,9.76) .. (7.2,9.6)
			 ..controls (7.107,9.44) and (7.213,8.933) .. (7.4,8.8)
			 ..controls (7.587,8.667) and (7.853,8.907) .. (8.2,9.)
			 ..controls (8.547,9.093) and (8.973,9.04) .. (9.2,9.2)
			 ..controls (9.427,9.36) and (9.453,9.733) .. (9.2,9.8)
			 ..controls (8.947,9.867) and (8.413,9.627) .. (8.,9.6);
\draw [very thick] (4.5,7.) ..controls (4.55,7.033) and (4.6,7.067) .. (4.7,7.1)
			 ..controls (4.8,7.133) and (4.95,7.167) .. (5.1,7.2);
\draw [very thick, ->] (5.8,7.2) ..controls (5.983,7.167) and (6.167,7.133) .. (6.3,7.1)
			 ..controls (6.433,7.067) and (6.517,7.033) .. (6.6,7.);
\draw [->, ultra thick] (4.2,9.2) ..controls (4.342,9.343) and (4.483,9.486) .. (4.7,9.6)
			 ..controls (4.917,9.714) and (5.208,9.8) .. (5.5,9.8)
			 ..controls (5.792,9.8) and (6.083,9.714) .. (6.3,9.6)
			 ..controls (6.517,9.486) and (6.658,9.343) .. (6.8,9.2);
\draw [->, very thick] (2.8,8.4) -- (2.8,7.7);
\draw [->, very thick] (8.4,8.4) -- (8.4,7.7);
\draw [] (5.2,7.) -- (5.4,7.2);
\draw [] (5.4,7.2) -- (5.6,7.4);
\draw [] (5.2,7.4) -- (5.4,7.2);
\draw [] (5.4,7.2) -- (5.6,7.);
\node [] at (5.4,6.7) {\scriptsize $\text{anomaly}$};
\node [] at (2.6,9.4) {\scriptsize $\text{G.S.}$};
\node [] at (8.2,9.4) {\scriptsize $\text{G.S.}$};
\node [] at (2.4,10.2) {\scriptsize $\text{QD}_{Z_2\times Z_2}$};
\node [] at (8.4,10.2) {\scriptsize $\text{QD}_{Z_2\times Z_2}$};
\node [] at (5.4,10.2) {\scriptsize $\text{TQD}_{Z_2\times Z_2}$};
\node [] at (9.4,8.8) {\scriptsize $\beta^1$};
\node [] at (3.8,8.5) {\scriptsize $\beta^0=1$};
\node [] at (2.4,7.5) {\scriptsize $\text{TFT}^{t=0}$};
\node [] at (8.2,7.5) {\scriptsize $\text{TFT}^{t=1}$};
\end{tikzpicture}
}

\newcommand{\ChargeDiagram}{
\begin{tikzpicture}[scale=0.9]
\draw [->-] (1.6,9.2) -- (2.2,9.8);
\draw [->-] (2.2,9.8) -- (1.6,10.4);
\draw [->-] (4.,9.2) -- (3.4,9.8);
\draw [->-] (3.4,9.8) -- (4.,10.4);
\draw [snake it] (2.2,9.8) -- (2.8,9.8);
\draw [snake it] (2.8,9.8) -- (3.4,9.8);
\node [] at (2.1,9.2) {\scriptsize $j_1$};
\node [] at (2.1,10.4) {\scriptsize $j'_1$};
\node [] at (3.5,9.2) {\scriptsize $j_2$};
\node [] at (3.5,10.4) {\scriptsize $j'_2$};
\node [] at (2.4,9.6) {\scriptsize $s$};
\node [] at (3.2,9.6) {\scriptsize $s'$};
\node [] at (2.8,10.2) {\scriptsize $\tilde{h}_{s,s'}$};
\draw [gray, fill] (2.8,9.8) circle [x radius=0.2, y radius=0.2];
\end{tikzpicture}
}

\newcommand{\ChargeDiagramAA}{
\begin{tikzpicture}[scale=0.9]
\draw [->-] (1.6,9.2) -- (2.2,9.8);
\draw [->-] (2.2,9.8) -- (1.6,10.4);
\draw [->-] (4.,9.2) -- (3.4,9.8);
\draw [->-] (3.4,9.8) -- (4.,10.4);
\draw [snake it] (2.2,9.8) -- (3.4,9.8);
\node [] at (2.1,9.2) {\scriptsize $j_1$};
\node [] at (2.1,10.4) {\scriptsize $j'_1$};
\node [] at (3.5,9.2) {\scriptsize $j_2$};
\node [] at (3.5,10.4) {\scriptsize $j'_2$};
\node [] at (2.8,10.) {\scriptsize $\tilde{\zeta}_s$};
\node [] at (2.8,9.6) {\scriptsize $s$};
\end{tikzpicture}
}

\newcommand{\CochainSpace}{
\begin{tikzpicture}[scale=1]
\draw [->, very thick] (3.4,9.7) ..controls (3.375,9.733) and (3.35,9.767) .. (3.4,9.8)
			 ..controls (3.45,9.833) and (3.575,9.867) .. (3.7,9.9);
\draw [very thick] (2.6,8.6) ..controls (2.62,8.54) and (2.64,8.48) .. (2.8,8.6)
			 ..controls (2.96,8.72) and (3.26,9.02) .. (3.4,9.2)
			 ..controls (3.54,9.38) and (3.52,9.44) .. (3.5,9.5);
\draw [very thick] (3.7,9.9) ..controls (3.783,9.867) and (3.867,9.833) .. (3.9,9.8)
			 ..controls (3.933,9.767) and (3.917,9.733) .. (3.9,9.7);
\draw [very thick, ->] (3.8,9.5) ..controls (3.76,9.453) and (3.72,9.407) .. (3.7,9.3)
			 ..controls (3.68,9.193) and (3.68,9.027) .. (3.7,8.9)
			 ..controls (3.72,8.773) and (3.76,8.687) .. (3.8,8.6);
\draw [very thick] (3.8,8.6) ..controls (3.633,8.66) and (3.467,8.72) .. (3.3,8.8)
			 ..controls (3.133,8.88) and (2.967,8.98) .. (2.9,9.1)
			 ..controls (2.833,9.22) and (2.867,9.36) .. (2.9,9.5);
\draw [very thick, ->] (2.9,9.7) ..controls (2.917,9.725) and (2.933,9.75) .. (2.9,9.8)
			 ..controls (2.867,9.85) and (2.783,9.925) .. (2.7,10.);
\draw [very thick] (2.7,10.) ..controls (2.617,9.975) and (2.533,9.95) .. (2.5,9.9)
			 ..controls (2.467,9.85) and (2.483,9.775) .. (2.5,9.7);
\draw [very thick, ->] (2.6,9.5) ..controls (2.65,9.375) and (2.7,9.25) .. (2.7,9.1)
			 ..controls (2.7,8.95) and (2.65,8.775) .. (2.6,8.6);
\draw [] (2.4,10.2) -- (4.8,10.2);
\draw [] (4.8,10.2) -- (4.,9.6);
\draw [] (1.6,9.6) -- (2.4,10.2);
\draw [] (4.8,9.) -- (4.,8.4);
\draw [] (4.,8.4) -- (1.6,8.4);
\draw [] (1.6,8.4) -- (2.4,9.);
\draw [] (1.6,9.6) -- (4.,9.6);
\draw [] (2.4,9.) -- (2.588,9.);
\draw [] (2.788,9.) -- (2.9,9.);
\draw [] (3.1,9.) -- (3.164,9.);
\draw [] (3.364,9.) -- (3.572,9.);
\draw [] (3.772,9.) -- (4.8,9.);
\draw [fill] (2.6,8.6) circle [x radius=0.08, y radius=0.08];
\draw [fill] (3.7,9.9) circle [x radius=0.08, y radius=0.08];
\draw [fill] (2.7,10.) circle [x radius=0.08, y radius=0.08];
\draw [fill] (3.8,8.6) circle [x radius=0.08, y radius=0.08];
\node [] at (2.3,8.6) {\scriptsize $1$};
\node [] at (4.1,10.) {\scriptsize $\beta$};
\node [] at (4.2,8.8) {\scriptsize $\beta^2$};
\node [] at (3.1,10.) {\scriptsize $\beta^3$};
\node [] at (3.4,10.4) {\scriptsize $\text{nontrivial cohomology class} [-1]$};
\node [] at (3.4,8.2) {\scriptsize $\text{trivial cohomology class} [1]$};
\end{tikzpicture}
}

\newcommand{\FigAnyonAB}{
\begin{tikzpicture}[scale=0.8]
\draw [lightgray, fill] (1.6,9.8) ..controls (1.867,9.876) and (2.134,9.952) .. (2.4,10.)
			 ..controls (2.666,10.048) and (2.932,10.068) .. (3.2,10.1)
			 ..controls (3.468,10.132) and (3.74,10.178) .. (4.,10.2)
			 ..controls (4.26,10.222) and (4.508,10.221) .. (4.8,10.2)
			 ..controls (5.092,10.179) and (5.426,10.137) .. (5.6,10.1)
			 ..controls (5.774,10.063) and (5.787,10.032) .. (5.8,10.);
\path [lightgray, fill] (1.6,9.4) -- (1.6,9.8) -- (4.,10.) -- (5.8,10.) -- (5.8,9.4) -- cycle;
\draw [ultra thick] (1.6,9.8) ..controls (1.867,9.876) and (2.134,9.952) .. (2.4,10.)
			 ..controls (2.666,10.048) and (2.932,10.068) .. (3.2,10.1)
			 ..controls (3.468,10.132) and (3.74,10.178) .. (4.,10.2)
			 ..controls (4.26,10.222) and (4.508,10.221) .. (4.8,10.2)
			 ..controls (5.092,10.179) and (5.426,10.137) .. (5.6,10.1)
			 ..controls (5.774,10.063) and (5.787,10.032) .. (5.8,10.);
\draw [fill] (1.6,9.8) -- (1.9,9.5);
\draw [fill] (1.9,9.5) -- (2.4,10.);
\draw [fill] (2.4,10.) -- (2.8,9.5);
\draw [fill] (2.8,9.5) -- (3.2,10.1);
\draw [fill] (3.7,9.6) -- (4.,10.2);
\draw [fill] (4.,10.2) -- (4.5,9.6);
\draw [fill] (4.5,9.6) -- (4.8,10.2);
\draw [fill] (4.8,10.2) -- (5.,9.7);
\draw [fill] (5.,9.7) -- (5.6,10.1);
\draw [] (3.7,9.6) -- (4.5,9.6);
\draw [] (4.5,9.6) -- (5.,9.7);
\draw [] (5.,9.7) -- (5.8,9.6);
\draw [] (3.2,10.1) -- (3.7,9.6);
\draw [] (3.7,9.6) -- (2.8,9.5);
\draw [] (2.8,9.5) -- (1.9,9.5);
\draw [] (4.5,9.6) -- (4.4,9.4);
\draw [] (4.5,9.6) -- (4.6,9.4);
\draw [] (5.,9.7) -- (4.8,9.4);
\draw [] (5.,9.7) -- (5.2,9.4);
\draw [] (3.7,9.6) -- (3.6,9.4);
\draw [] (3.7,9.6) -- (3.8,9.4);
\draw [] (2.8,9.5) -- (2.7,9.4);
\draw [] (2.8,9.5) -- (2.9,9.4);
\draw [] (1.6,9.6) -- (1.9,9.5);
\draw [] (1.9,9.5) -- (1.9,9.4);
\draw [fill, outer color=white,white] (1.7,10.) circle [x radius=0.2, y radius=0.2];
\draw [fill, outer color=white,white] (2.5,10.2) circle [x radius=0.2, y radius=0.2];
\draw [fill, outer color=white,white] (3.4,10.3) circle [x radius=0.2, y radius=0.2];
\draw [fill, outer color=white,white] (4.,10.4) circle [x radius=0.2, y radius=0.2];
\draw [fill, outer color=white,white] (5.,10.4) circle [x radius=0.2, y radius=0.2];
\draw [fill, outer color=white,white] (5.8,10.2) circle [x radius=0.2, y radius=0.2];
\end{tikzpicture}
}

\newcommand{\FigAnyonAC}{
\begin{tikzpicture}[scale=0.8]
\draw [lightgray, fill] (1.6,9.7) ..controls (1.867,9.776) and (2.134,9.852) .. (2.4,9.9)
			 ..controls (2.666,9.948) and (2.932,9.968) .. (3.2,10.)
			 ..controls (3.468,10.032) and (3.739,10.078) .. (4.,10.1)
			 ..controls (4.261,10.122) and (4.513,10.121) .. (4.8,10.1)
			 ..controls (5.087,10.079) and (5.411,10.037) .. (5.6,10.)
			 ..controls (5.789,9.963) and (5.845,9.932) .. (5.9,9.9);
\path [lightgray, fill] (1.6,9.3) -- (1.6,9.7) -- (4.,9.9) -- (5.9,9.9) -- (5.9,9.3) -- cycle;
\draw [ultra thick] (1.6,9.7) ..controls (1.867,9.776) and (2.134,9.852) .. (2.4,9.9)
			 ..controls (2.666,9.948) and (2.932,9.968) .. (3.2,10.)
			 ..controls (3.468,10.032) and (3.739,10.078) .. (4.,10.1)
			 ..controls (4.261,10.122) and (4.513,10.121) .. (4.8,10.1)
			 ..controls (5.087,10.079) and (5.411,10.037) .. (5.6,10.)
			 ..controls (5.789,9.963) and (5.845,9.932) .. (5.9,9.9);
\draw [fill] (1.6,9.7) -- (1.9,9.4);
\draw [fill] (1.9,9.4) -- (2.4,9.9);
\draw [fill] (2.4,9.9) -- (2.8,9.4);
\draw [fill] (2.8,9.4) -- (3.2,10.);
\draw [fill] (3.7,9.5) -- (4.,10.1);
\draw [fill] (4.,10.1) -- (4.5,9.5);
\draw [fill] (4.5,9.5) -- (4.8,10.1);
\draw [fill] (4.8,10.1) -- (5.,9.6);
\draw [fill] (5.,9.6) -- (5.6,10.);
\draw [] (3.7,9.5) -- (4.5,9.5);
\draw [] (4.5,9.5) -- (5.,9.6);
\draw [] (5.,9.6) -- (5.9,9.5);
\draw [] (3.2,10.) -- (3.7,9.5);
\draw [] (3.7,9.5) -- (2.8,9.4);
\draw [] (2.8,9.4) -- (1.9,9.4);
\draw [] (4.5,9.5) -- (4.4,9.3);
\draw [] (4.5,9.5) -- (4.6,9.3);
\draw [] (5.,9.6) -- (4.8,9.3);
\draw [] (5.,9.6) -- (5.2,9.3);
\draw [] (3.7,9.5) -- (3.6,9.3);
\draw [] (3.7,9.5) -- (3.8,9.3);
\draw [] (2.8,9.4) -- (2.7,9.3);
\draw [] (2.8,9.4) -- (2.9,9.3);
\draw [] (1.6,9.5) -- (1.9,9.4);
\draw [] (1.9,9.4) -- (1.9,9.3);
\draw [fill, outer color=white,white] (1.8,9.9) circle [x radius=0.2, y radius=0.2];
\draw [fill, outer color=white,white] (2.4,10.2) circle [x radius=0.2, y radius=0.2];
\draw [fill, outer color=white,white] (3.4,10.) circle [x radius=0.2, y radius=0.2];
\draw [fill, outer color=white,white] (4.,10.4) circle [x radius=0.2, y radius=0.2];
\draw [fill, outer color=white,white] (5.,10.2) circle [x radius=0.2, y radius=0.2];
\draw [fill, outer color=white,white] (4.4,9.8) circle [x radius=0.2, y radius=0.2];
\draw [fill, outer color=white,white] (4.,9.5) circle [x radius=0.2, y radius=0.2];
\draw [fill, outer color=white,white] (3.,9.5) circle [x radius=0.2, y radius=0.2];
\draw [fill, outer color=white,white] (2.2,9.5) circle [x radius=0.2, y radius=0.2];
\draw [fill, outer color=white,white] (5.,9.7) circle [x radius=0.2, y radius=0.2];
\draw [fill, outer color=white,white] (5.4,10.2) circle [x radius=0.2, y radius=0.2];
\draw [fill, outer color=white,white] (5.8,9.6) circle [x radius=0.2, y radius=0.2];
\end{tikzpicture}
}

\newcommand{\TriangulationN}{
\begin{tikzpicture}[scale=1]
\draw [->-] (3.8,9.3) ..controls (3.816,9.16) and (3.831,9.02) .. (3.8,8.9)
			 ..controls (3.769,8.78) and (3.691,8.68) .. (3.6,8.6)
			 ..controls (3.509,8.52) and (3.404,8.46) .. (3.3,8.4);
\draw [->-] (3.3,8.4) ..controls (3.178,8.353) and (3.056,8.307) .. (2.9,8.3)
			 ..controls (2.744,8.293) and (2.556,8.327) .. (2.4,8.4)
			 ..controls (2.244,8.473) and (2.122,8.587) .. (2.,8.7);
\draw [->-] (2.7,10.4) ..controls (2.8,10.408) and (2.9,10.417) .. (3.,10.4)
			 ..controls (3.1,10.383) and (3.2,10.342) .. (3.3,10.3);
\draw [->-] (3.3,10.3) ..controls (3.404,10.251) and (3.509,10.202) .. (3.6,10.1)
			 ..controls (3.691,9.998) and (3.769,9.842) .. (3.8,9.7)
			 ..controls (3.831,9.558) and (3.816,9.429) .. (3.8,9.3);
\draw [->-] (2.,8.7) ..controls (1.917,8.833) and (1.833,8.967) .. (1.8,9.1)
			 ..controls (1.767,9.233) and (1.783,9.367) .. (1.8,9.5);
\draw [dashed] (1.8,9.5) ..controls (1.82,9.587) and (1.84,9.673) .. (1.9,9.8)
			 ..controls (1.96,9.927) and (2.06,10.093) .. (2.2,10.2)
			 ..controls (2.34,10.307) and (2.52,10.353) .. (2.7,10.4);
\draw [->-] (2.,8.7) -- (2.8,9.3);
\node [] at (2.24,9.21) {\scriptsize $a_{1}$};
\draw [->-] (3.3,10.3) -- (2.8,9.3);
\draw [->-] (3.3,8.4) -- (2.8,9.3);
\node [] at (3.3,8.99) {\scriptsize $a_{2}$};
\draw [->-] (3.8,9.3) -- (2.8,9.3);
\node [] at (3.3,9.52) {\scriptsize $a_{3}$};
\node [] at (2.6,8.6) {\scriptsize $b{1}$};
\node [] at (4.,8.7) {\scriptsize $b{2}$};
\node [] at (4.,10.) {\scriptsize $b{3}$};
\node [] at (1.6,9.1) {\scriptsize $b_{N}$};
\end{tikzpicture}
}

\newcommand{\TriangulationNAA}{
\begin{tikzpicture}[scale=1]
\draw [->-] (3.6,9.3) ..controls (3.616,9.16) and (3.631,9.02) .. (3.6,8.9)
			 ..controls (3.569,8.78) and (3.491,8.68) .. (3.4,8.6)
			 ..controls (3.309,8.52) and (3.204,8.46) .. (3.1,8.4);
\draw [->-] (3.1,8.4) ..controls (2.978,8.353) and (2.856,8.307) .. (2.7,8.3)
			 ..controls (2.544,8.293) and (2.356,8.327) .. (2.2,8.4)
			 ..controls (2.044,8.473) and (1.922,8.587) .. (1.8,8.7);
\draw [->-] (2.5,10.4) ..controls (2.6,10.408) and (2.7,10.417) .. (2.8,10.4)
			 ..controls (2.9,10.383) and (3.,10.342) .. (3.1,10.3);
\draw [->-] (3.1,10.3) ..controls (3.204,10.251) and (3.309,10.202) .. (3.4,10.1)
			 ..controls (3.491,9.998) and (3.569,9.842) .. (3.6,9.7)
			 ..controls (3.631,9.558) and (3.616,9.429) .. (3.6,9.3);
\draw [->-] (1.8,8.7) ..controls (1.717,8.833) and (1.633,8.967) .. (1.6,9.1)
			 ..controls (1.567,9.233) and (1.583,9.367) .. (1.6,9.5);
\draw [dashed] (1.6,9.5) ..controls (1.62,9.587) and (1.64,9.673) .. (1.7,9.8)
			 ..controls (1.76,9.927) and (1.86,10.093) .. (2.,10.2)
			 ..controls (2.14,10.307) and (2.32,10.353) .. (2.5,10.4);
\draw [->-] (1.8,8.7) -- (2.6,9.3);
\draw [->-] (2.7,8.3) -- (2.6,9.3);
\draw [->-] (3.6,8.9) -- (2.6,9.3);
\draw [->-] (3.6,9.7) -- (2.6,9.3);
\draw [->-] (3.1,10.3) -- (2.6,9.3);
\draw [fill, outer color=white,white] (1.8,8.7) circle [x radius=0.2, y radius=0.2];
\draw [fill, outer color=white,white] (2.7,8.3) circle [x radius=0.2, y radius=0.2];
\draw [fill, outer color=white,white] (3.6,8.9) circle [x radius=0.2, y radius=0.2];
\draw [fill, outer color=white,white] (3.6,9.7) circle [x radius=0.2, y radius=0.2];
\draw [fill, outer color=white,white] (3.1,10.3) circle [x radius=0.2, y radius=0.2];
\node [] at (1.7,8.3) {\scriptsize $j_{1}$};
\node [] at (3.3,8.3) {\scriptsize $j_{2}$};
\node [] at (3.8,8.5) {\scriptsize $j_{3}$};
\node [] at (3.8,10.) {\scriptsize $j_{4}$};
\node [] at (3.5,10.4) {\scriptsize $j_{5}$};
\end{tikzpicture}
}

\begin{document}

\title{From effective Hamiltonian to anomaly inflow in topological orders with boundaries}
\author{Yuting Hu}
\email{yuting.phys@gmail.com}
\affiliation{Department of Physics and Center for Field Theory and Particle Physics, Fudan University, Shanghai 200433, China}
\author{Yidun Wan}
\email{ydwan@fudan.edu.cn}
\affiliation{Department of Physics and Center for Field Theory and Particle Physics, Fudan University, Shanghai 200433, China}
\affiliation{Collaborative Innovation Center of Advanced Microstructures, Nanjing 210093, China}
\author{Yong-Shi Wu}
\email{yswu@fudan.edu.cn}
\affiliation{State Key Laboratory of Surface Physics, Fudan University, Shanghai 200433, China}
\affiliation{Department of Physics and Center for Field Theory and Particle Physics, Fudan University, Shanghai 200433, China}
\affiliation{Collaborative Innovation Center of Advanced Microstructures, Nanjing 210093, China}
\affiliation{Department of Physics and Astronomy, University of Utah, Salt Lake City, Utah, 84112, U.S.A.}
\date{\today}

\begin{abstract}
Whether two boundary conditions of a two-dimensional topological order can be continuously connected without a phase transition in between remains a challenging question. We tackle this challenge by constructing an effective Hamiltonian, describing anyon interaction, that realizes such a continuous deformation. At any point along the deformation, the model remains a fixed point
model describing a gapped topological order with gapped boundaries. That the deformation retains the gap is due to the anomaly cancelation between the boundary and bulk. Such anomaly inflow is quantitatively studied using our effective Hamiltonian. We apply our method of effective Hamiltonian to the extended twisted quantum double model with boundaries (constructed by two of us in Ref.\cite{Bullivant2017}). We show that for a given gauge group $G$ and a three-cocycle in $H^3[G,U(1)]$ in the bulk, any two gapped boundaries for a fixed subgroup $K\subseteq G$ on the boundary can be continuously connected via an effective Hamiltonian. Our results can be straightforwardly generalized to the extended Levin-Wen model with boundaries (constructed by two of us in Ref.\cite{Hu2017a}).  

\end{abstract}
\maketitle

\noindent{\textit{Introduction}}:
Topologically ordered matter systems have greatly expanded our knowledge of matter phases\cite{Wen1989,Wen1989a,Wen1990a,Wen1990c,Kitaev2003a,Levin2004,Kitaev2006,Chen2012a,Levin2012,Hung2012,Hu2012,Hu2012a,Mesaros2011,Lin2014,Kong2014}, may potentially be used as quantum memories\cite{Dennis2002}, and realize topological quantum computation\cite{Kitaev2003a,Freedman2003,Stern2006,Nayak2008}. Among all the factors that hinder the physical applicability of topological orders, a crucial one is that topological orders have been studied mostly for closed two-dimensional systems, whereas experimentally realizable materials mostly have boundaries. When a topological order is placed on an open surface, a boundary is subject to certain gapped boundary condition on which the topological order remains well-defined. It remains however a challenge whether two apparently different gapped boundary conditions of a topological order are physically equivalent. There has been a few constructions of boundary Hamiltonians of topological orders \cite{Beigi2011,Kitaev2012,Cong2016a,Cong2017a,Wang2017}, which are nevertheless for either restricted cases or in the language of categories. Very recently, in Ref.\cite{Hu2017,Hu2017a,Bullivant2017}, we have systematically constructed the boundary Hamiltonians of the Levin-Wen\cite{Levin2004} and the twisted quantum double models (TQD)\cite{Hu2012a} using solely the microscopic degrees of freedom of the models. This allows us to tackle the challenge aforementioned. 

To do so, we adopt the extended Hamiltonian constructed in Ref.\cite{Bullivant2017} for the TQD model with boundaries. Without loss of generality, we consider the case with only one boundary, namely a disk. Such an extended TQD model $H^{G,\alpha}_{K,\beta}$ defined by a finite gauge group $G$ and a $3$-cocycle $\alpha\in H^3[G,U(1)]$ in the bulk, and a subgroup $K\subseteq G$ and a $2$-cocycle $\beta\in H^2[K,U(1)]$ on the boundary. We prove that $H^{G,\alpha}_{K,\beta}\sim H^{G,\alpha}_{K,1} $ for all $\beta$ by showing that $H^{G,\alpha}_{K,\beta}$ is connected to $H^{G,\alpha}_{K,1} $ via an  continuous passage that retains the gap. Such an  continuous passage can be understood as a unitary transformation relating the Hilbert spaces before and after the  continuous passage. It is proposed in Ref.\cite{WenTensorCat2004} that two topological orders on a closed surface are equivalent if and only if they are related by finite steps of local unitary transformations. In the case with boundaries, however, we find  that the unitary transformation associated with an  continuous passage is local in the bulk but nonlocal on the boundary. That the system remains gapped throughout the entire  continuous passage is due to the anomaly inflow from the boundary to bulk, which is corroborated by the nonlocal unitary transformation on the boundary spectrum. We derive an emergent (effective) Hamiltonian $\tilde H_{}$ that realizes the  continuous passage $\exp{i\tilde H t}$ (parameterized by $t$) between the two models $H^{G,\alpha}_{K,\beta}$ and $H^{G,\alpha}_{K,1} $. Using this emergent Hamiltonian, we quantitatively study the nonlocal unitary transformation and the anomaly inflow.

Our results hold for the extended TQD model with any Abelian finite group $G$. We accompany our derivation with an explicit example---the extended TQD model with gauge group $G=\Z_2\times \Z_2$. Our results are generic, which can also apply to the extended Levin-Wen model with boundaries systematically constructed in Ref.\cite{Hu2017,Hu2017a}.
\\

\begin{figure}[h!]
\centering
\subfigure[]{\includegraphics[scale=0.6]{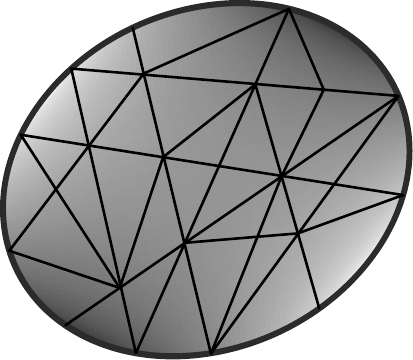}}
\subfigure[]{\TriangulationN}
\subfigure[]{\TriangulationNAA}
\caption{(a) An arbitrary triangulation of a disk. (b) A reduced triangulation of the disk. The labels refer to group elements. (c) Same graph as (b) with however labels referring to quasiparticles, i.e., representations of the group elements.}
\label{fig:disk}
\end{figure}

\noindent{\textit{Extended TQD on a disk}}:
We place the TQD model with gauge group $G$ on a graph that triangulates a disk, as in Fig. \ref{fig:disk}. That the model is a low-energy fixed point effective theory leads to the topological invariance of the model\cite{Hu2012a,Bullivant2017}, such that the initial arbitrary graph can be reduced by the Pachner moves\cite{Pachner1978,Hu2012a,Bullivant2017} into the simple form in Fig. \ref{fig:disk}(b). The reduced graph consists of $N+1$ vertices (one bulk vertex $0$ and $N$ boundary vertices) and $2N$ edges ($N$ bulk edges $a_1$, through $a_N$ and $N$ boundary edges $b_1$ through $b_N$). 
The bulk edge degrees of freedom $a_n$'s take value in The boundary edge degrees of freedom $b_n$'s take value in certain subgroup $K\subseteq G$. The Hamiltonian of the model on the reduced graph reads
\begin{equation}\label{eq:ThreeHamiltonian}
H^{G,\alpha}_{K,\beta}=-\sum_{v=0}^N A_v-\sum_{p=1}^N B_p. 
\end{equation} 
where $A_v$ are the vertex operators acting on vertices $v$, and $B_p$ are the plaquette operators acting on the plaquettes.  One can check that the operators in Hamiltonian \eqref{eq:ThreeHamiltonian} are commuting projection operators, and the ground-state space are invariant under topology-preserving graph mutations (i.e., Pachner moves).
The matrix elements of these operators are combinations of a $3$-cocycle $\alpha\in H^3[G,U(1)]$ and an $\alpha$-dependent $2$-cocycle $\beta\in H^2[K,U(1)]$ satisfying the Frobenius condition 
\begin{equation}\label{eq:AlphaDependentBeta}
\alpha\delta\beta=1,
\end{equation}
where $\delta$ denotes the $3$-coboundary operator.
(Mathematically, this defines $\beta$ as a Frobenius algebra in the category $Vect_G^{\alpha}$.)
\\

\noindent\textit{Deformation class of extended TQD models}: Given an extended TQD model $H^{G,\alpha_0}_{K,\beta_0}$, we can construct a deformation class of extended TQD models $\{H^{G,\alpha^t}_{K,\beta^t}\}$ for a continuous parameter $t$, with
\begin{equation}\label{eq:deformDef}
\alpha^t=\alpha_0(\delta\xi^{t})^{-1},
\beta^t=\beta_0\xi^t.
\end{equation}
where $\xi^t$ is an arbitrary $U(1)$-valued function (i.e., a $2$-cochain) with initial condition $\xi^{t=0}=1$, i.e., $\alpha^{t=0}=\alpha_0$ and $\beta^{t=0}=\beta_0$.

We check that $H^{G,\alpha^t}_{K,\beta^t}$ is a well-defined extended TQD model at any $t$. To see this, one can verify that
\begin{equation}\label{eq:topologicalInvariance}
\delta\alpha^t=1,
\quad
\alpha^t\delta(\beta^t)=1,
\end{equation}
By the first condition, $\alpha^t$ is a $3$-cocycle on $G$, which defines the bulk TQD Hamiltonian. By the second condition, $\beta^t$ is $\alpha^t$-dependent $2$-cocycle on $K$, which defines the boundary Hamiltonian. Hence $H^{G,\alpha^t}_{K,\beta^t}$ is an extended TQD model. During the deformation, the energy spectrum of the system remains the same. That is, there is no level crossing and thus no phase transition.

To better understand the above general approach and the physical consequences, let us work on an explicit example hereafter. 
\\

\begin{figure}[h!]
\centering
\subfigure[]{\CochainSpace\label{fig:CochainSpace}}
\subfigure[]{\includegraphics[scale=0.62]{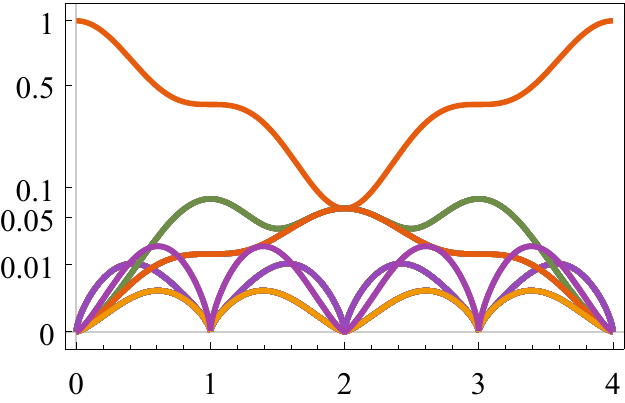}\label{fig:deformAmplitude}}
\caption{(a) Parameter space for the Hamiltonian deformation. Hamiltonian $H$ depends on $\beta ^t$ that goes along a path in the $2$-cochain space. The path starts at $\beta=1$ at $t=0$, goes through a nontrivial $2$-cocycle $\beta$ at $t=1$, then $\beta^2\in[1]$ (but not identical to 1)  at $t=2$, and another nontrivial 2-cocycle $\beta^3$ at $t=3$, and goes back to $\beta^4=1$ at $t=4$. (b) Variation of $|C^{j_1,\dots,j_N}_t|^2$ along the deformation passage for $N=3$.}
\end{figure}

\noindent \textit{Example $G=K=\Z_2\x\Z_2$}: 
This is the simplest example for a nontrivial continuous deformation. The precise form of the matrix elements of the operators in the Hamiltonian \eqref{eq:ThreeHamiltonian} in this case are recorded in Appendix \ref{appd:AvBp}. Since $H^2[\Z_2\x\Z_2,U(1)]=\Z_2$, the $2$-cocycles are grouped into two equivalence classes $[1]$ and $[-1]$. We then  restrict to the case with $\alpha_0=1$ and $\beta_0=1$ at $t=0$, such that the initial extended TQD model reduces to the $\Z_2\x\Z_2$ Kitaev QD model on a disk with a trivial boundary condition. We can then construct the continuous deformation \eqref{eq:deformealpha} with the one-parameter family 
\begin{equation}\label{eq:deformBetaMatrix}
\beta^t(a,b)=
\begin{pmatrix}
\begin{array}{cccc}
1 & 1 & 1 & 1 \\
1 & 1 & e^{\frac{1}{2} \ii \pi  t} & e^{-\frac{1}{2} \ii \pi  t} \\
1 & e^{-\frac{1}{2}\ii\pi t} & 1 & e^{\frac{1}{2}\ii\pi t} \\
1 & e^{\frac{1}{2}\ii\pi t} & e^{-\frac{1}{2} \ii \pi  t} & 1 \\
\end{array}
\end{pmatrix},
\end{equation}
which is indexed by $a,b=00,01, 10, 11\in\Z_2\x\Z_2$ and satisfies $\beta^t(a,a^{-1}b)=1/\beta^t(b,b^{-1}a)$.  Correspondingly, we set $\alpha^t=\delta\beta^{-t}$. We recognize that
\begin{equation}\label{eq:CocycleTable}
\beta^t({a,b})=\beta^{t+2}(a,b)^{-1}=
\begin{pmatrix}
\begin{array}{cccc}
1 & 1 & 1 & 1 \\
1 & 1 & i & -i \\
1 & -i & 1 & i \\
1 & i & -i & 1 \\
\end{array}
\end{pmatrix},\ t\in 4\Z+1,
\end{equation}
and $\beta^t(a,b)=\beta^{t+2}(a,b)^{-1}=1$ if $t\in 4\Z$. Consequently, a closed deformation loop forms for $t\in[0,4]$. See Fig. \ref{fig:CochainSpace}. The $\alpha^t=1$ if and only if $t$ is an integer. In this figure, one can see that only the four big dots correspond to extended QD models, whereas any other point along the deformation loop corresponds to an extended TQD model. That is, the deformation between two extended QD models would have to go into the space of extended TQD models.

\noindent \textit{Effective 1+1D Hamiltonian of interacting anyons}: To understand the  continuous deformation, let us begin with the ground-state  wavefunction $\Phi(t)$ on the disk, with an explicit  $t$ dependence,
\begin{align}\label{eq:GSt}
\Phi_t=\frac{1}{|G|^N}\prod_{n=1}^{N}\beta^t(a_n,a_n^{-1}a_{n+1})\delta_{b_n,a_n^{-1}a_{n+1}},
\end{align}
where and hereafter we let $N+1=1$. This is the $t$-deformation of the wavefunction obtained in Ref.\cite{Bullivant2017}.

The excitations are characterized by topological quasiparticles, or, anyons, in the bulk and on the boundary. There are two types of quasiparticles: charges identified by $A_{v=0}$ in the bulk and by $A_{v}$ on all boundary vertices $1$ through $N$; and flux identified by $B_p$ on bulk triangles. By examining the ground state wavefunction, we see no flux will appear duration deformation for all $t$. Hence we consider excitations with only charges in the bulk and on the boundary. We first express a basis of excitations with charges $j_1$ through $j_N$ residing respectively at the $N$ vertices on the boundary as
\begin{align}\label{eq:eigenstatesBasis}
\Psi^{j_1\dots j_N}_{t}=\frac{1}{|G|^N}\prod_{n=1}^N\rho^{j_n}(a_n)\Phi_t,
\end{align}
where $\rho^j(a)$ is an irreducible representation of $Z_2\times \Z_2$. See Fig. \ref{fig:disk}(c). For $j_n=j_Lj_R\in\{00,01,10,11\}$ with $j_L,j_R\in\{0,1\}$, $\rho^j(a)$ takes the form
\begin{equation}\label{eq:RepresentationZ2Z2}
\rho^j(a)=\exp[\pi\ii (j_La_L+j_Ra_R)].
\end{equation}
Here, the charge $00$ is the trivial one or vacuum. 
In such a basis  $\Psi^{j_1\dots j_N}$, however, there is also a charge $-\sum_{n=1}^N j_n$ residing at vertex $0$ in the bulk, due to a global constraint that the total charge of the system is null.

The basis states $\Psi^{j_1\dots j_N}_t$ are always the energy eigenstates at time $t$ but not at any other $t'\neq t$.
This deformation in fact defines a one-parameter family of continuous (unitary) transformation on the anyon bases at different $t$ values, which quantifies how anyons recombine and/or shuffle during the deformation. In the following, we will rewrite the ground state $\Phi(t)$ at $t$ as a linear combination of excitations at $t=0$. Namely, using Eq. \eqref{eq:GSt} and \eqref{eq:eigenstatesBasis}, we decompose $\Phi_t$ as
\begin{equation}\label{eq:decomposePhi}
\ket{\Phi_t}=\sum_{j_1,\dots, j_N}C^{j_1\dots j_N}_t\ket{ \Psi^{j_1\dots j_N}_{t=0}}
\end{equation}
with
\begin{align}\label{eq:decomposeCoeff}
C^{j_1\dots j_N}_t
=\frac{1}{|G|^N}
\sum_{a_1\dots a_N}
\prod_{n=1}^{N}\rho^{-j_n}(a_n)\beta^t(a_n,a_n^{-1}a_{n+1}).
\end{align}
Using
\begin{equation}\label{eq:ThetaToBeta}
\beta^t(a,b)=\exp[\ii h(a,b)t],
\end{equation}
Eq. \eqref{eq:decomposePhi} can be differentiated as
\begin{equation}\label{eq:DifferentiatePhi}
-\ii \partial_t\ket{\Phi_t}=\tilde{H}\ket{\Phi_t},
\end{equation}
where
\begin{equation}\label{eq:emergentHam}
\tilde{H}=\sum_{n=1}^{N}\tilde{H}_n
\end{equation}
with
\begin{equation}\label{eq:emergentHamN}
\tilde{H}_n\ket{\Psi^{\dots j_nj_{n+1}\dots}_t}
=\sum_{j'_nj'_{n+1}}\tilde{h}_{j'_n-j_n,j'_{n+1}-j_{n+1}}\ket{\Psi^{\dots j'_nj'_{n+1}\dots}_t},
\end{equation}
where
\begin{align}\label{eq:Gamma}
\tilde{h}_{s,s'}
=\frac{1}{|G|^2}\sum_{ab}\rho^{-s}(a)\rho^{-s'}(b) h(a,a^{-1}b),
\end{align}
where $s=j'_n-j_n$ and $s'=j'_{n+1}-j_{n+1}$. In the equations  above, the anyon charges in $\dots$ of $\Psi$ remain intact. The interaction $\tilde{h}$ quantifies the exchange of anyon charges between two neighboring anyons.  
\begin{figure}
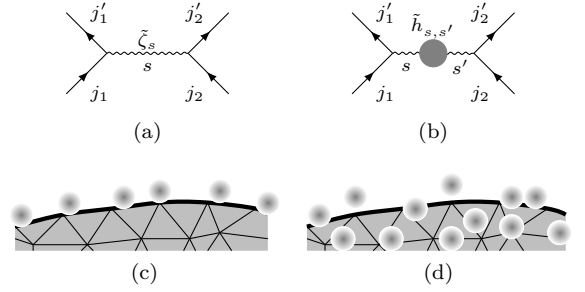

\centering
\subfigure[]{\ChargeDiagramAA}
\qquad \quad \quad
\subfigure[]{\ChargeDiagram}
\subfigure[]{\FigAnyonAB} 
\subfigure[]{\FigAnyonAC}

\caption{The effective anyon interaction due to (a) $\tilde h_{s,s'}$ in Eq. \eqref{eq:Gamma} that does not conserve anyon charge because $s\neq s'$ generally, (b) $\tilde\zeta_s$ in Eq. \eqref{eq:ZetaTilde} that conserves anyon charge. (c) The unconservation is due to bulk-boundary anyon charge exchange. (d) The conservation is because the anyon charges shuffle and recombines on the boundary only.}
\label{fig:chargeflow}
\end{figure}

In our example, $\tilde{h}$ reads explicitly as a matrix
\begin{equation}\label{eq:hmatrix}
\tilde{h}
=
\frac{\pi}{8}
\begin{pmatrix}
\begin{array}{cccc}
0 & 0 & 0 & 0 \\
0 & 0 & 1 & -1 \\
0 & -1 & 0 & 1 \\
0 & 1 & -1 & 0 \\
\end{array}
\end{pmatrix}
\end{equation}
with matrix indexed by $s,s'=00,01,10$ and $11$.

Consider a  continuous deformation $H(t)$. The parameter can be viewed as a virtual time, while Eq. \eqref{eq:emergentHam} defines an emergent Hamiltonian describing the interactions of the anyons on boundaries. Such a Hamiltonian determines the adiabatic evolution of the ground state $\Phi_t$. The $t$-dependence of the probability amplitudes $|C^{j_1,j_2,\dots,j_N}|^2$ is illustrated in Fig. \ref{fig:deformAmplitude}.

In the anyon basis, we introduce $4\times 4$ matrix $\tau_{j}^x$ defined by Pauli matrices
\begin{equation}\label{eq:smatrix}
\tau^{00}=1\otimes 1,
\tau^{01}=1\otimes \sigma^x,
\tau^{10}=\sigma^x\otimes 1,
\tau^{11}=\sigma^x\otimes \sigma^x.
\end{equation}

Then the effective Hamiltonian becomes a spin chain
\begin{equation}\label{eq:HamiltonianinTau}
\tilde{H}_n=\sum_{ss'}\tilde{h}_{s,s'}\tau_n^s\tau_{n+1}^{s'}.
\end{equation}

\noindent{\textit{Charge Conservation}}: 
When $\beta^t=\delta\gamma$, i.e., $\beta$ can be expressed as a 2-coboundary
\begin{equation}\label{eq:betaGamma}
\beta(a,b)=\gamma(a)\gamma(b)/\gamma(ab).
\end{equation}
Anyon interactions preserves the total charge. To see this, let $\gamma(a)=\exp\{\ii t \zeta(a)\}$, and define the corresponding Forier transformation
\begin{equation}\label{eq:ZetaTilde}
\tilde{\zeta}_s=\frac{1}{|G|}\sum_a\rho^{-s}(a)\zeta(a).
\end{equation}
We express $\theta$ as
\begin{equation}\label{eq:ThetaZeta}
\theta(a,a^{-1}b)=\zeta(a)+\zeta(a^{-1}b)-\zeta(b)
\end{equation}
The two terms $\zeta(a)$ and $\zeta(b)$ are canceled by the sum in Eq. \eqref{eq:emergentHam}. The remaining term in emergent Hamiltonian is given by
\begin{equation}\label{eq:GammaGamma}
\tilde{h}_{s,s'}=\delta_{-s,s'}\tilde{\zeta_{s'}},
\end{equation}        
where the delta function implies the total charge conservation during the anyon interaction.  This is illustrated in Fig. \ref{fig:chargeflow}(a). Consequently, the boundary anyons only recombine and shuffle on the boundary. See Fig. \ref{fig:chargeflow}(c). The effective spin-chain Hamiltonian now reads
\begin{equation}\label{eq:GammaSpinChain}
\tilde{H}_n=\sum_{s}\tilde{\zeta}_s\tau^{-s}_{n}\tau^{s}_{n+1}.
\end{equation}

For example, in the deformation \eqref{eq:deformBetaMatrix}, we can define a new path to deform $H(t=0)$ to $H(t=2)$, with $\tilde{\zeta}$ being
\begin{equation}\label{eq:ZetaExample}
\tilde{\zeta}=\frac{\pi}{4}(1,-1,-1,1).
\end{equation}

In general, however, $\beta^t\neq\delta\gamma$, such that the interaction $\tilde h_{s,s'}$ \eqref{eq:Gamma} does not conserve the anyon charge, namely, $j_1'+j_2'\neq j_1+j_2$ because $s\neq -s'$ in general, as in Fig. \ref{fig:chargeflow}(b). Had the boundary been a stand-alone $(1+1)$-D system, this charge unconservation would cause anomaly. Nonetheless, in our $(2+1)$-D system, the excessive anyon charges does not disappear but leaks into the bulk and cancel the anyon charges in the bulk, as sketched in Fig. \ref{fig:chargeflow}(c). Such charge unconversation implies anomaly cancelation via anomaly inflow, which we now explain and quantify. 
\\

\begin{figure}[h!]
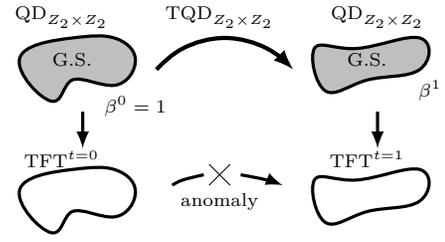

\centering
\AnomalyPic
\caption{Pure $(1+1)$-D theories defined by $\beta_1$ and $\beta_2$ cannot be continuously connected. This anomaly is canceled when $\beta_1$ and $\beta_2$ define two boundaries of the same $(2+1)$-D theory.}
\label{fig:anomalyInflow}
\end{figure}   

\noindent{\textit{Anomaly inflow}}: Consider the two extended QD theories in the upper two corners of Fig. \ref{fig:anomalyInflow}, where the bulk is restricted to ground states. There exists two stand-along $(1+1)$-D theories, denoted by $\text{TFT}^{t=0}$ and $\text{TFT}^{t=1}$ in the lower two corners of Fig. \ref{fig:anomalyInflow}. Coupling these two  $(1+1)$-D theories to a pure gauge theory in the bulk (determined by $\alpha^{t=0}=\alpha^{t=1}=1$) results in the the two extended QD theories as just mentioned.

Now consider a deformation from $\text{TFT}^{t=0}$ to $\text{TFT}^{t=1}$, not coupled to a bulk. As stand-along (1+1)-D theories, $\text{TFT}^{t=0}$ and $\text{TFT}^{t=1}$ belong to different phases, characterized by two inequivalent 2nd-cohomology classes $[1]$ and $[-1]$ respectively. Hence there must be a phase transition during the deformation. Upon a transition point, the system is gapless, and the corresponding $(1+1)$-D theory is anomalous.

Such an anomaly is a gauge anomaly for the following reason. The anyon charges are gauge charges (with $\Z_2\times\Z_2$ viewed as the gauge group). The violation of conservation of boundary anyon charges in the extended QD models implies the violation of gauge invariance of the (1+1)-D TFT theories. Hence the anomaly is a gauge anomaly. The conservation of the total anyon charges in the entire system (bulk plus boundary) implies that the gauge anomaly is canceled by the bulk. Therefore, the inflow of anyon charges from boundary to bulk quantitatively characterizes the gauge anomaly inflow.

\begin{figure}[h!]
\centering
\includegraphics[scale=0.5]{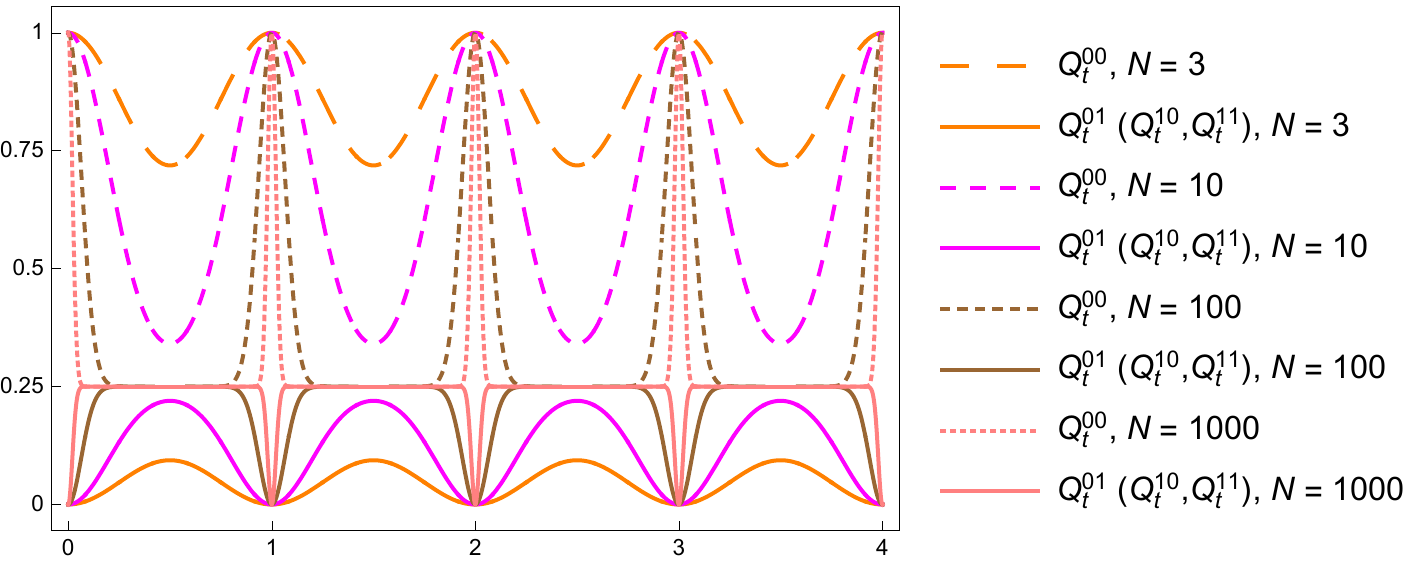}
\caption{The accumulative total anyon-charge exchange.}
\label{fig:AnomalyCancelation}
\end{figure}

We define the total anyon-charge exchange between the boundary and bulk accumulatively from $t=0$ to $t$ to be
\begin{align}\label{eq:movingCharge}
Q^j_t=&\sum_{j_1\dots j_N}|C^{j_1\dots j_N}_t|^2\delta_{j_1+\dots +j_N,j}
\nonumber\\
=&\frac{1}{|G|^{N+1}}
\sum_{xa_1\dots a_N}\rho^j(x)
\prod_{n}^N\frac{\beta^t(a_n,a_n^{-1}a_{n+1})}{\beta^t(xa_n,a_n^{-1}a_{n+1})}.
\end{align}
We compute $Q^j_t$ for $N=3$, $10$, $100$, and $1000$, using Eq. \eqref{eq:deformBetaMatrix}. See Fig. \ref{fig:AnomalyCancelation}. We can see that in the thermodynamic limit $N\rightarrow\infty$, $Q^j_t\rightarrow 1/4$ for all $j$, i.e., evenly distributed. More importantly, $Q^j_t\equiv 0$ at integer $t$ for all $j\neq 00$, which quantitatively demonstrates the anomaly cancellation by anomaly inflow, as the bulk at integer $t$ is described by the same pure gauge theory.
\\

\begin{acknowledgements}
YH and YDW thank J\"urgen Fuchs, Ling-Yan Hung, Christopher Schweigert, and Kenichi Shimizu for very helpful discussions. YDW is also supported by the Shanghai Pujiang Program.  
\end{acknowledgements}

\begin{appendix}

\section{Vertex and plaquette operators}\label{appd:AvBp}     
Here we list the action of the vertex operators in the Hamiltonian \eqref{eq:ThreeHamiltonian} for the case with $N=3$ in Fig. \ref{fig:disk}(b). \begin{align}\label{eq:ThreeAZero}
&A_{v=0}\state{a_1 & a_2 & a_3}{b_1 & b_2 & b_3}\nonumber\\
=&\frac{1}{|G|}\sum_{h}\frac{\alpha(h,a_1,b_3^{-1})}{\alpha(h,a_1,b_1)\alpha(h,a_2,b_2)}
\state{ha_1 & ha_2 & ha_3}{b_1 & b_2 & b_3}
\end{align}
\begin{align}\label{eq:ThreeAOne}
&A_{v=1}\state{a_1 & a_2 & a_3}{b_1 & b_2 & b_3}
\nonumber\\
=&\frac{1}{|G|}\sum_{h}\frac{\alpha(a_1h^{-1},h,b_1)\beta(h,b_1)}{\alpha(a_1h^{-1},h,b_3^{-1})\beta(b_1h^{-1},h)}
\state{a_1h^{-1} & a_2 & a_3}{hb_1 & b_2 & b_3h^{-1}}
\end{align}
\begin{align}\label{eq:ThreeATwo}
&A_{v=2}\state{a_1 & a_2 & a_3}{b_1 & b_2 & b_3}
\nonumber\\
=&\frac{1}{|G|}\sum_{h}\frac{\alpha(a_2h^{-1},h,b_2)\beta(h,b_2)}{\alpha(a_1,b_2h^{-1},h)\beta(b_1h^{-1},h)}
\state{a_1 & a_2h^{-1} & a_3}{b_1h^{-1} & hb_2 & b_3}
\end{align}
\begin{align}\label{eq:ThreeAThree}
&A_{v=3}\state{a_1 & a_2 & a_3}{b_1 & b_2 & b_3}\nonumber\\
=&\frac{1}{|G|}\sum_{h}\frac{\alpha(a_1,b_3^{-1}h^{-1},h)\beta(b_3^{-1}h^{-1},h)}{\alpha(a_2,b_2h^{-1},h)\beta(b_2h^{-1},h)}
\state{a_1 & a_2 & a_3h^{-1}}{b_1 & b_2h^{-1} & hb_3}.
\end{align}
Here $\state{a_1 & a_2 & a_3}{b_1 & b_2 & b_3}$ is a shorthand notation for a state on the reduced graph in Fig. \ref{fig:disk}(b) for $N=3$. The plaquette operators $B_p$ is defined on triangles. On a triangle $p$, $B_p=1$ if the product of the three group elements along the three edges of the triangle clockwise is equal to the identity element of the group, and $B_p=0$ otherwise.

\section{Symmetry condition}\label{appd:symm}
The 2-cocycles used in this our computation has symmetry
\begin{align}\label{eq:symmetryCondition}
&\beta(a,b)=\beta(b,b^{-1}a^{-1})=\beta(b^{-1},a^{-1})^{-1}\nonumber\\
&\beta({00,a})=\beta({a,00)}=\beta({a,a)}=1.
\end{align}
This implies
\begin{equation}\label{eq:symm}
\tilde{h}_{s,s'}=-\tilde{h}_{-s',-s}.
\end{equation}
\end{appendix}        

\bibliographystyle{apsrev}
\bibliography{StringNet}

\end{document}